\documentclass[a4paper]{spie} 
\usepackage[T1,OT2]{fontenc}
\usepackage[latin1]{inputenc}
\usepackage[english]{babel}
\usepackage{graphicx}
\usepackage{subfigure}
\usepackage{float}
\usepackage{amsmath}
\usepackage[nice]{nicefrac}
\usepackage{geometry}
\usepackage[nottoc, notlof, notlot]{tocbibind}
\usepackage{url}
\usepackage{textcomp}
\usepackage{verbatim}
\usepackage{aeguill}
\usepackage{fancyhdr}
\usepackage{etex}
\usepackage{m-pictex,m-ch-en}
\usepackage{pstricks,pstricks-add,pst-math,pst-xkey}
\usepackage{minitoc}
\usepackage{multirow}
\usepackage{amstext}

\DeclareSymbolFont{cyrletters}{OT2}{wncyr}{m}{n}
\DeclareMathSymbol{\Sha}{\mathalpha}{cyrletters}{"58}

\newcommand{\nvec}[1]{\text{\boldmath$#1$}}

\title{$C_{n}^{2}$ profile from Shack-Hartmann data with CO-SLIDAR data processing}

\author{Clélia Robert, Juliette Voyez, Nicolas
  Védrenne, Laurent Mugnier
  \skiplinehalf
  \scriptsize{ONERA, DOTA/HRA, 29 Avenue de la Division Leclerc, 92322
    Châtillon, France}}

\authorinfo{E-mail: clelia.robert@onera.fr}

\begin{document}

\maketitle

\begin{abstract}

  $C_{n}^{2}$ profile monitoring usually makes use of wavefront slope 
  correlations or of scintillation pattern correlations. Wavefront slope
  correlations provide sensitivity to layers close to the receiving plane. In
  addition, scintillation correlations allow a better sensitivity to high
  turbulence layers. Wavefront slope and scintillation
  correlations are therefore complementary. Slopes and scintillation being
  recorded simultaneously with a Shack-Hartmann wavefront sensor (SHWFS), we
  propose here to exploit their correlation to retrieve the $C_{n}^{2}$ 
  profile. The measurement method named COupled SLodar scIDAR (CO-SLIDAR)\cite{Vedrenne:07}
  uses correlations of SHWFS data from two separated stars. A
  maximum-likelihood method is developed to estimate precisely the positions and
  intensities corresponding to each SHWFS spot, which are used as inputs for
  CO-SLIDAR. First results are presented using SHWFS real data from a binary star.

\end{abstract}


\section{Introduction}

New Adaptive Optics (AO) systems, such as MCAO, GLAO, LTAO, have been conceived to
optimize wavefront correction on different fields of view (FOV), but their
efficiency very much depends on knowledge of the turbulence vertical
distribution. A more precise determination of the turbulence strength profile
$C_{n}^{2}$ is therefore needed to improve their performances. A $C_{n}^{2}$
profile can be obtained indirectly from meteorological parameters, but it is more
usually measured directly by optical means. These means depend on the number of
sources employed and type of data involved. In Generalized SCIDAR
(Scintillation Detection and Ranging)\cite{1998PASP..110...86F}, the $C_{n}^{2}$
profile is retrieved from correlation of the scintillation pattern produced by a
binary star in a pupil plane. SLODAR (Slope Detection and Ranging)\cite{2002MNRAS.337..103W} uses
instead wavefront slope correlations
measured on a binary star with a SHWFS.

We propose a new approach for $C_{n}^{2}$ profile measurement named CO-SLIDAR;
using a SHWFS, it means both slope and intensity data can be fruitfully
utilized. With CO-SLIDAR, slope correlations recorded on two separated stars
deliver low-altitude layer sensitivity as a SLODAR. In addition, scintillation correlations and
correlations between slopes and scintillation (which is referred further to
coupling) deliver high-altitude layer sensitivity. With a
limited pupil size and in a single instrument, CO-SLIDAR conjugates the
advantages of MASS (Multi-Aperture Scintillation
Sensor)\cite{2003MNRAS.343..891T} and DIMM (Differentiel Image Motion
Monitor)\cite{2007MNRAS.382.1268K}, possibly with better resolution.
CO-SLIDAR has been validated numerically\cite{Vedrenne:07}. Recently we tested
this method with a SHWFS on a single infrared source\cite{Vedrenne:10}. In
order to then quantify actual CO-SLIDAR performance on a double
star, we here test a new smart estimator that processes subapertures images to extract slopes and intensities.

In Section 2 we recall the analytical background of a $C_{n}^{2}$ profile measurement
based on exploitation of the correlations between SHWFS data: slope
correlations, scintillation correlations and their coupling. In Section 3, we
present the new estimator based on a maximum-likelihood criterion to measure
positions
and intensities precisely in a given subaperture. First results using real data from a
binary star are presented in Section 4. In Section 5, we sum up our
conclusions and open perspectives about CO-SLIDAR.

\section{Problem statement with SHWFS slope and scintillation correlations}

Given a star with position $\nvec{\alpha}$ in the FOV, a SHWFS
delivers a set of wavefront slopes and intensities per frame. The slope computed on
the $m^{th}$ subaperture focal image is a bi-dimensional vector
$\nvec{s}_m(\nvec{\alpha})$ with two components $s^k_m(\nvec{\alpha})$, along
the $k$ axis ($k \in \{x,y\}$). Star intensities, denoted
$i_m(\nvec{\alpha})$ and recorded in every sub-aperture $m$, lead to scintillation
index $\delta i_m(\nvec{\alpha})
=\frac{i_m(\nvec{\alpha})-o_m(\nvec{\alpha})}{o_m(\nvec{\alpha})}$ where
$o_m(\nvec{\alpha})$ is the time-averaged star intensity.

For two stars separated from $\nvec{\theta}$, correlations of SHWFS data are
empirically estimated from a finite number of frames. Slope
correlations $\overline{ s^k_m s^l_n}(\nvec{\theta}) $, scintillation index
correlations $\overline{\delta i_m \delta i_n}(\nvec{\theta})$ and their
coupling $\overline{s^k_m \delta i_n}(\nvec{\theta})$ are stacked in a
single dimension covariance vector $\nvec{C_{mes}}$. It relates directly to
$C_{n}^{2}$ in the problem statement as follows:
\begin{equation}
  \label{eq:Delta_H_CO_SLIDAR}
  \nvec{C_{mes}} = \mathcal{M} \nvec{C_n^2} + \nvec{C_d} + \nvec{u}
\end{equation}

where $\mathcal{M}$ is the interaction function; $\nvec{C_d}$ is the
covariance vector of the noises affecting slope and intensity measurements;
$\nvec{u}$ represents uncertainties on $\nvec{C_{mes}}$ due to the limited
number of frames. In Rytov regime, $\mathcal{M}$ is a linear operator $M$ and
depends on SHWFS geometry, statistical properties of the turbulence, star
separation, distance between sub-apertures and the set of discretized
altitudes\cite{2006JOSAA..23..613R}. The column
vectors of $M$ are formed by the concatenation of weighting functions, for
slope correlations, scintillation correlations and their coupling. They can
be seen as correlations induced by a turbulent layer at altitude h, for a
certain distance between subapertures. Layer
contribution to $\nvec{C_{mes}}$ have been discussed in a recent
publication\cite{Vedrenne:07} and therefore shall cover that material.

In processing our first real data with CO-SLIDAR (see Section 4), we do not implement
$\nvec{C_d}$, the covariance vector of noises affecting slope and
intensity measurements, that bias the wavefront slope and scintillation
correlation estimate. We do have data with a good signal to noise ratio
($SNR \simeq 100$) and the estimator to compute slopes and intensities is noise-free,
so $\nvec{C_d}$ is negligible. Nevertheless, if it were not the case, assumming the system is well calibrated,
it would be possible to completely determine $\nvec{C_d}$ and define a non-biased
estimation of the covariance vector,
$\nvec{\hat{C}_{mes}}=\nvec{C_{mes}}-\nvec{C_d}$. This would make it possible to rewrite the
direct problem:

\begin{equation}
  \label{eq:new_direct_problem}
  \nvec{\hat{C}_{mes}}= MC_{n}^{2}+\nvec{u}
\end{equation}

A sampled estimate of $C_{n}^{2}$, $\nvec{\tilde S}$, is retrieved from
the inversion of Eq. \ref{eq:new_direct_problem}, assuming that the
convergence noise $u$ is Gaussian.
For physical reasons $C_{n}^{2}$ is never negative, so $\nvec{\tilde S}$
minimizes the maximum likelihood criterion $J$ under positivity constraint:

\begin{equation}
  \label{eq:J_a_priori}
  J = (\nvec{\hat{C}_{mes}} - M \nvec{\tilde S})^T C_{conv}^{-1}
  (\nvec{\hat{C}_{mes}} - M \nvec{\tilde S})
\end{equation}

where $C_{conv}=\langle \nvec{u} \nvec{u}^T\rangle$ is the covariance matrix
of $\nvec{u}$, the convergence noise, due to the limited number of frames.

\section{A new estimator for positions and intensities in a SHWFS subaperture}

Usually, wavefront slopes are estimated by center of gravity (COG). However, here
we also need to measure scintillation indexes for each star in the SHWFS
subapertures. We therefore present a new algorithm, called Reconstarfield,
that has been developed to estimate positions and
intensities corresponding to each SHWFS spot precisely.
The principle of the estimator is to minimize a maximum likelihood criterion
between the star field image and its model, when the point spread
function (PSF) and number of stars is known. 
The minimized criterion $J^{'}$ is the following:

\begin{equation}
  \label{eq:criterion_reconstarfield}
 J^{'} \left(\{a_{n},x_{n},y_{n}\}_{n=1}^{N},b\right) =
  \sum_{p,q}w\left(p,q\right)\left|i\left(p,q\right)-i_{0}\left(p,q,\{a_{n},x_{n},y_{n}\}_{n=1}^{N},
b\right)\right|^{2} 
\end{equation}

where:
\begin{itemize}

\item
  $i_{0}\left(p,q,\{a_{n},x_{n},y_{n}\}_{n=1}^{N},b\right)
  =\left[\sum_{n=1}^{N}a_{n}h_{n}\left(x-x_{n},y-y_{n}\right)\right]_{\Sha}\left(p,q\right)+b$
  is the image model, considering the positions $\left(x_{n},y_{n}\right)$ and
  intensities $a_{n}$ of the $n$ stars. The index $n$ is included between $1$
  and $N$, $N$ being the number of stars in the field. $\Sha$  denotes the
  delta-comb representing the sampling operation. $b$ is the background level.

\item $h_{n}$ is the PSF of the spot $n$, taking into account both optical transfer and
  detector transfer functions.

\item $i\left(p,q\right)=i_{0}\left(p,q\right)+n\left(p,q\right)$ is the
  recorded image on the detector at pixel $\left(p,q\right)$, namely the sampled image with noise
  $n\left(p,q\right)$, which is assumed to be the sum of photon and detector
  noises, and can be approximated as Gaussian for the considered fluxes. 

\item $w\left(p,q\right)$ are the weights corresponding to the inverse of
  the noise variance. They can be inhomogeneous, but here we have
  supposed $w=1$, and $w=0$ can be used to eliminate the influence of bad pixels.

\end{itemize}

A raw estimation of the star positions, performed by a local COG, starts the
minimization. The background level $b$ and spot intensities $a_{n}$ are
estimated analytically for any given spot positions $\{x_{n},y_{n}\}$, which
eases and accelerates minimisation of $J^{'}$. The positions are then determined iteratively for the previous intensities with
Levenberg-Marquardt (LM) algorithm. 

Reconstarfield has been tested intensively in simulation for binary stars simulated with
Gaussian PSF and various separations including close binaries. Results close
to reality were obtained with greater precision than with the
COG, even in the presence of photon noise and narrow binaries. Realistic
SHWFS images were built including photon noise, using a set of $100$ images
featuring the characteristics of the real data described in Section 4, to
compute accuracy on the position evaluation in comparison with that of the COG, and
that of intensity retrieval. The precisions are respectively $\sigma_{pos}^{2}
\simeq 10^{-5}~pixel^{2}$ for the positions and $\sigma_{int}^{2} \simeq
10^{-5}$ in relative intensity.

\section{$C_{n}^{2}$ profile estimation using real data from a binary star}

We present here the data processing from a SHWFS installed on the Carlos S\'anchez
Telescope at the Teide Observatory (Canary Islands). 
It is a Cassegrain telescope, with a diameter of $D=1.5~m$ and a central obscuration
of $0.4D$. The source is a binary
star (BS5475), with a separation of $\theta =5.6~''$, with magnitudes of
$V=4.8$ and $5.9$. SHWFS images consist of $10 \times10$ subapertures, with a diameter
of $d=15~cm$. The data sequence contains $5000$ frames recorded with an iXon
EMCCD camera at $330~Hz$, so
the sequence duration is $15~s$. The wavefront is recorded  on a $120 \times
120~pixel$ detector, at a wavelength of $\lambda=0.6~\mu m$. We have assumed
monochromaticity of the source and in the
following we do not take into account the star spectrum's possible impact 
on the data. The subaperture field of view contains $12 \times 12~$pixels and
the image sampling is close to Shannon. We note that SHWFS altitude resolution  for a
SLODAR, expressed by $\delta h=\frac{d}{\theta}$, is insufficient, since it equals only $5~$km and
the maximum altitude, $H_{max}\simeq \frac{D}{\theta}$ is too high, about
$50~$km with such a star separation. We obtain this from the SHWFS geometry with
large subapertures of $15~$cm in diameter, and from the narrow binary
separation. This leads to refined signal processing, presented below, and
to optimized SHWFS design and experimentation to be proposed in the future.

\begin{figure}[!h]
  \begin{center}
    \begin{tabular}{cc}
      \includegraphics[width=6cm]{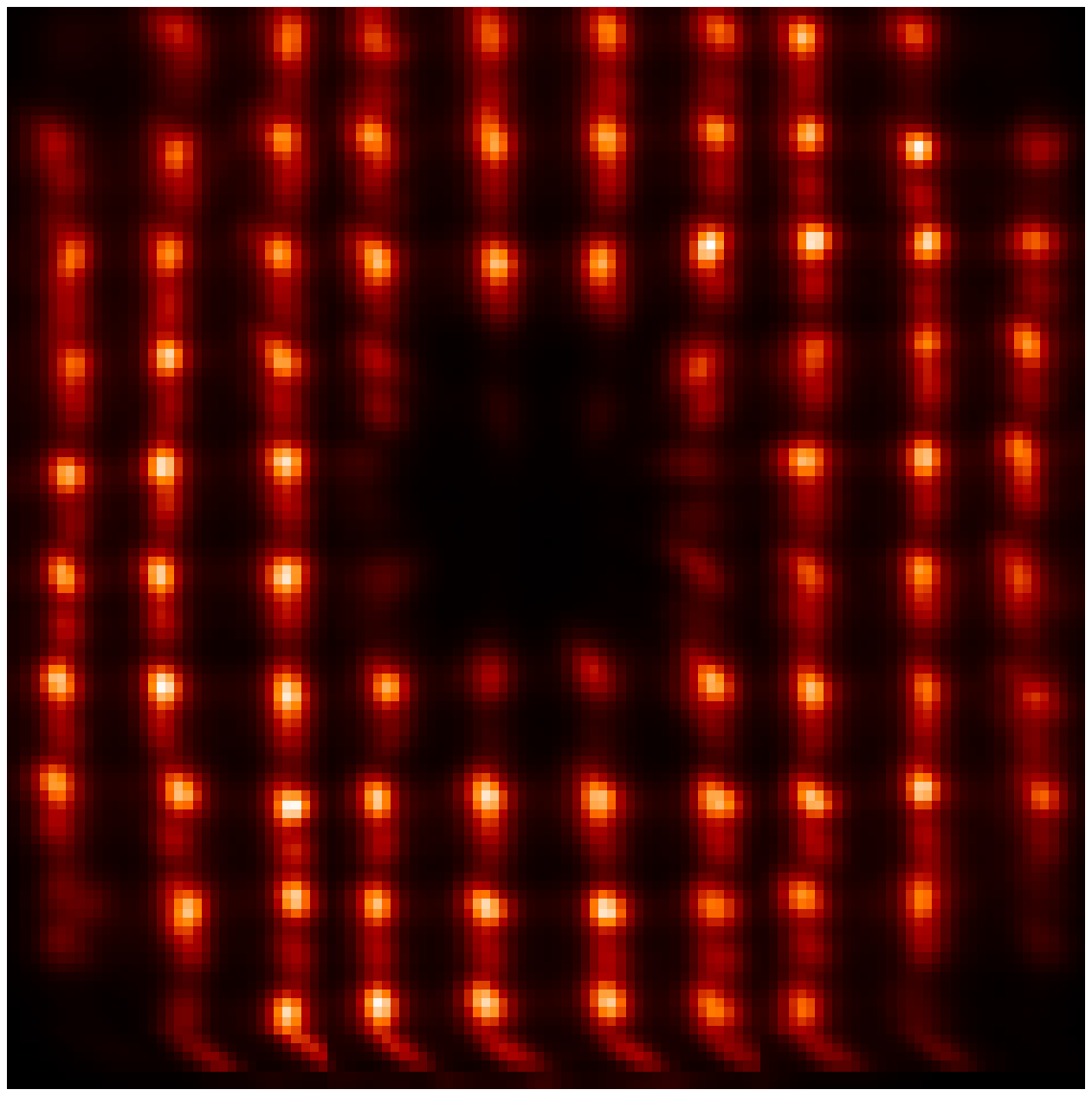} &
      \includegraphics[width=6cm]{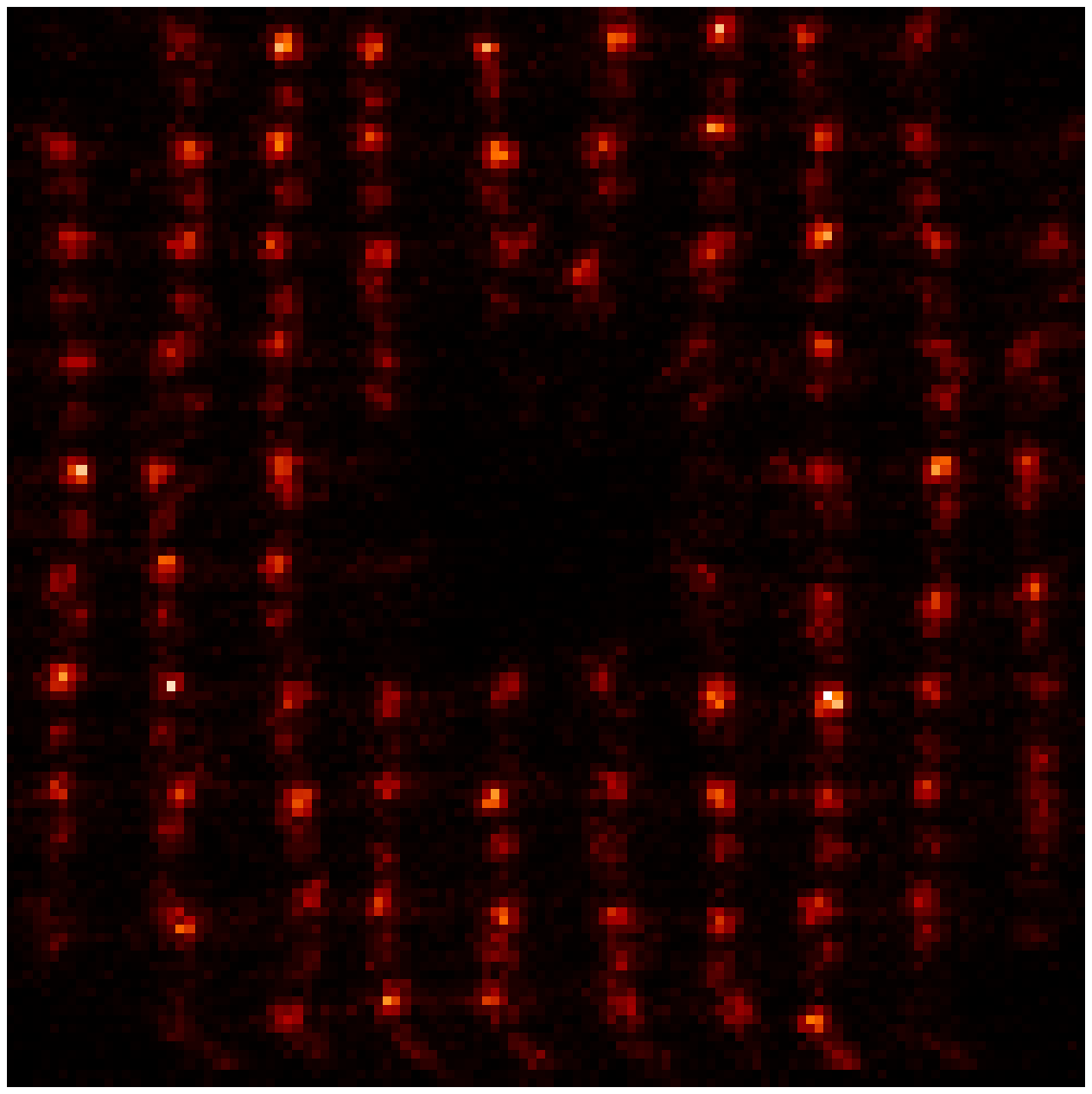}
    \end{tabular}
      \caption{SHWFS long-exposure image (on the left) and short-exposure
        image (on the right).\label{fig:image_shack}}
  \end{center}
\end{figure}

Fig. \ref{fig:image_shack} shows the SHWFS long-exposure image and an example
of a short-exposure image.
As we can see in Fig \ref{fig:image_shack}, we cannot use all
subapertures, because of the central obscuration and vignetted subapertures at
each corner of the images. Only $76$ subapertures remain for subaperture image
processing. We compute positions $\left(x_{n},y_{n}\right)$ and intensities ($i_{m}
\left( \alpha \right)$) for each
subaperture, for each star, using Reconstarfield. Slopes ($s_{m} \left( \alpha
\right)$) are obtained by substracting the
time-averaged position in each direction, for each star.
Scintillation indexes are calculated by using $\delta i_m(\nvec{\alpha})
=\frac{i_m(\nvec{\alpha})-o_m(\nvec{\alpha})}{o_m(\nvec{\alpha})}$, where
$o_m(\nvec{\alpha})$ is the time-averaged star intensity. The step of
extracting slopes and scintillation indexes being done, we 
cross-correlate them to feed $\nvec{\hat{C}_{mes}}$.

\begin{figure}[!h]
  \begin{center}
    \begin{tabular}{c}
      \includegraphics[width=10cm]{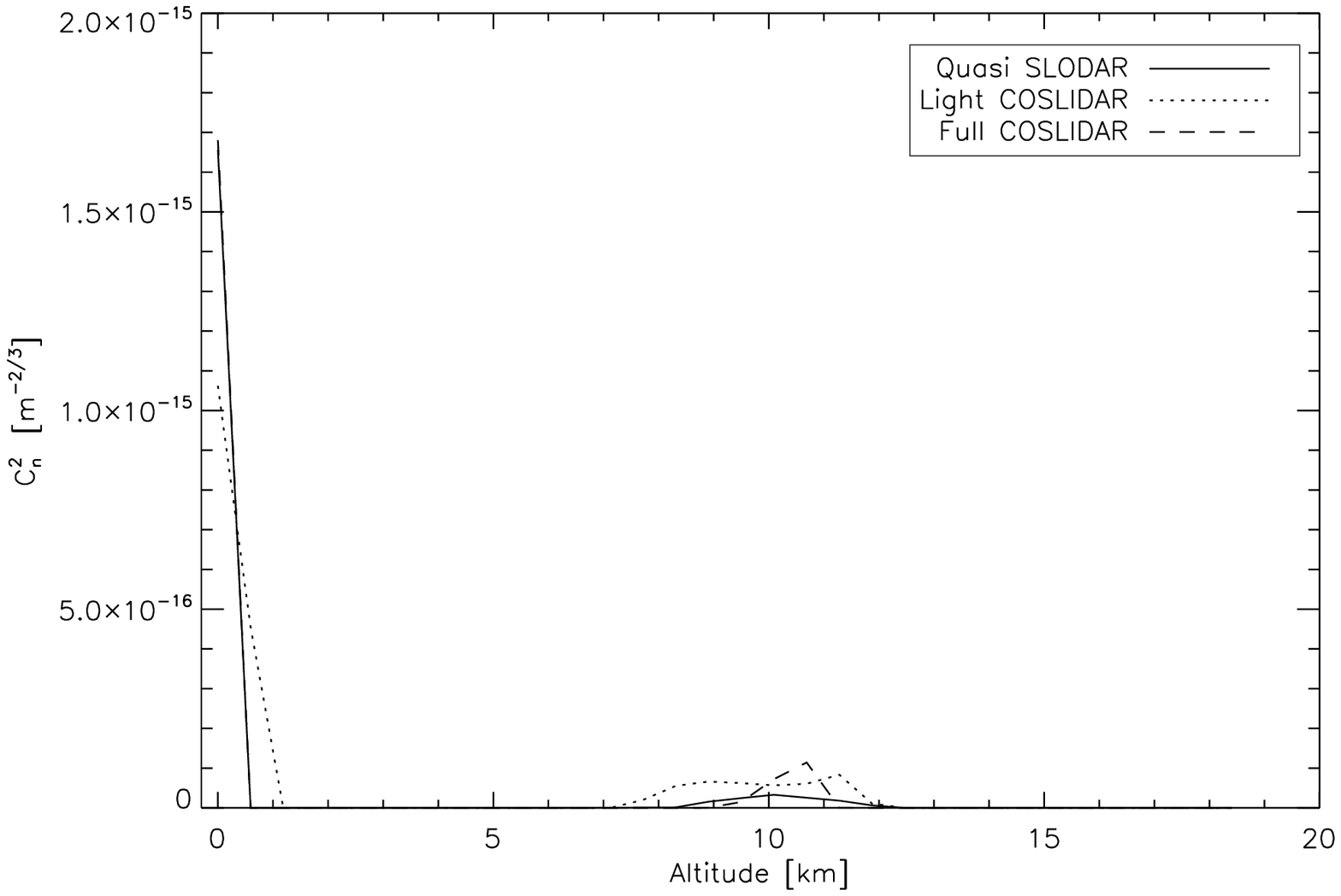} \\
      \includegraphics[width=10cm]{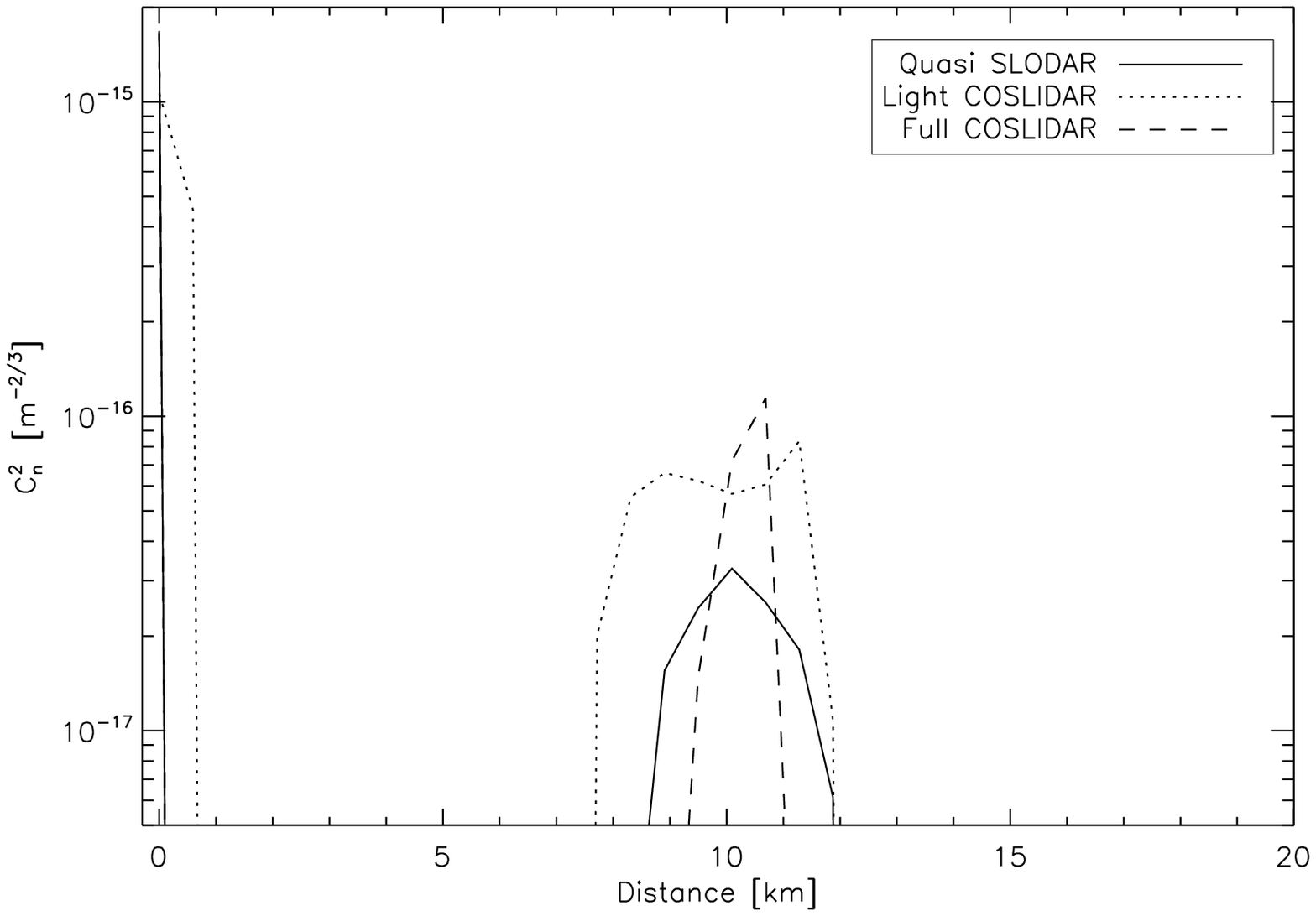} \\
    \end{tabular} 
    \caption{$C_{n}^{2}$ profile results from inversion using CO-SLIDAR data
      processing. Top: normal scaling on $y$ axis; bottom: log scaling on $y$ axis.
      \label{fig:Cn2_profiles}}
  \end{center}
\end{figure}

$C_{n}^{2}$ profile is retrieved from slope correlations, scintillation
correlations and their coupling, by minimizing the $J$ criterion given by Eq.
\ref{eq:J_a_priori}. The iterative method is an adaptive step gradient
descent. The positivity constraint is implemented by projection and we do not
use any regularization function. We first only use the slope
correlations to process the data in a ``Quasi SLODAR'' configuration, then add the scintillation
correlations. This data processing is known as Light CO-SLIDAR. Last,
we perform a complete CO-SLIDAR analysis, adding the coupling, and what is
known as 
Full CO-SLIDAR. The retrieved $C_{n}^{2}$ profiles corresponding to each
approach are presented in Fig. \ref{fig:Cn2_profiles} as a function of the altitude.

We retrieve a set of $32$ values of $C_{n}^{2}$ across $20~$km of atmosphere, proceeding
with super-resolution. Basically we would have retrieved only $10$ values with such a
SHWFS geometry. All types of processing detect a strong turbulent layer 
within the first kilometer. The Quasi SLODAR configuration shows a weak layer at
$10~$km. With Light CO-SLIDAR, the high-altitude layer is also detected at
$10~$km and is stronger and larger. The Full CO-SLIDAR method shows a similar layer at
the same altitude, but it is thinner. The three methods are
in good congruence since they recover layers at the same altitude. Further,
this $10-$km layer is present if we vary the number of layers reconstructed
between $10$, $20$ and $32$, or if we use the scintillation signal only. 

Using the Full CO-SLIDAR $C_{n}^{2}$ profile we derive some turbulence parameters. The
first one is Fried parameter $r_{0}$ defined by:

\begin{equation}
\label{eq:r0}
r_{0}=\left[0,423\left(\dfrac{2\pi}{\lambda}\right)^{2}\int_{0}^{H_{max}}C_{n}^{2}(h)dh\right]^{-\nicefrac{3}{5}}
\end{equation}

We compute $r_{0}$ by integrating the restored $C_{n}^{2}$ along the line of sight.
Another method is to use the slope variance of the most brillant star in each
subaperture for the whole sequence. Both techniques give
$r_{0} \simeq 9~cm$. The $C_{n}^{2}$ profile also allows to derive the isoplanatic
angle $\theta_{0}$ (Eq. \ref{eq:theta0}) and scintillation rate $\sigma_{scint}^{2}$ (Eq. \ref{eq:scintillation_rate}):

\begin{equation}
\label{eq:theta0}
\theta_{0}=\left[2.91\left(\dfrac{2\pi}{\lambda}\right)^{2}\int_{0}^{H_{max}}C_{n}^{2}(h)h^{\nicefrac{5}{3}}dh\right]^{-\nicefrac{3}{5}}
\end{equation}

\begin{equation}
\label{eq:scintillation_rate}
\sigma_{scint}^{2}= 4 \times 0.56 \left(\dfrac{2\pi}{\lambda}\right)^{\nicefrac{7}{6}}\int_{0}^{H_{max}}C_{n}^{2}(h)h^{\nicefrac{5}{6}}dh
\end{equation}

Numerical application of these equations gives $\theta_{0} \simeq 2.2~''$ and $\sigma_{scint}^{2} \simeq
0.09$. The values of $r_{0}$, $\theta_{0}$ and $\sigma_{scint}^{2}$ are
compared with those of a publication presenting synchronized SCIDAR
profiles\cite{2009MNRAS.397.1633G} to the data used in this paper. We get
a much smaller $r_{0}$, quite surprising for an astronomical
site, and comparable values of $\theta_{0}$ and $\sigma_{scint}^{2}$. The
$C_{n}^{2}$ profiles involved in computation of
turbulence parameters may explain the differences. In any event, the differential
scintillation is detectable since the value of $\sigma_{scint}^{2}$, typical
of an astronomical site, is measurable with Reconstarfield and since the
binary star separation is larger than the isoplanatic angle $\theta_{0}$.

\section{Conclusion and perspectives}

We have proposed a new approach for $C_{n}^{2}$ profile measurements with a
SHWFS, using the information provided by both slope and intensity data.
Testing of this concept has begun on experimental data. We have presented here
the first results from retrieving the $C_{n}^{2}$ profile with CO-SLIDAR from
SHWFS data recorded at the Teide Observatory. A new estimator was employed to
compute both positions and intensities in SHWFS subaperture images. The
$C_{n}^{2}$ profile is consistent with those of astronomical sites, featuring
a strong layer in the first kilometer of altitude and a weak layer at higher
altitude (about $10~$km). We plan an experience with a better-adapted SHWFS, to
retrieve a $C_{n}^{2}$ profile across the first $20~$km of atmosphere with a
$1~$km altitude resolution. This could be done with larger star separation and
smaller subapertures.

Additional work should be performed to quantify the number of
frames needed to achieve highest accuracy and to take into account detection
noise. An \textit{a posteriori} maximum-likelihood criterion should be
implemented for the inverse problem, to take a given
profile into account (\textit{i.e.} regularization). Outer-scale influence on the accuracy
of CO-SLIDAR should be investigated. Determination of wind profile and perhaps
outer-scale should also be studied.

\section{Acknowledgements}

This work is being performed in connection with a PhD thesis supported by
ONERA, the French Aerospace Lab, and the French Direction Générale de
l'Armement (DGA). The authors are very
grateful to B. Garcia-Lorenzo and A. Rodriguez for sharing the SHWFS data
recorded at the Teide Observatory and for their contribution in presenting
their instrument's information. The authors also thank  J.-M. Conan,
T. Fusco and V. Michau for fruitful exchanges they have had with them.


\end{document}